\documentclass[twocolumn,showpacs,preprintnumbers,amsmath,amssymb]{revtex4}

\usepackage{graphicx}
\usepackage{dcolumn}
\usepackage{bm}
\usepackage{subfigure}
\usepackage{bm}

\begin{document}

\preprint{SNN/2004-1}
\newcommand{\be}{\begin{equation}}
\newcommand{\ee}{\end{equation}}
\newcommand{\bc}{\begin{center}}
\newcommand{\ec}{\end{center}}
\newcommand{\bea}{\begin{eqnarray}}
\newcommand{\eea}{\end{eqnarray}}
\newcommand{\beaa}{\begin{eqnarray*}}
\newcommand{\eeaa}{\end{eqnarray*}}
\newcommand{\del}[3]{\left#1 #3 \right#2}
\newcommand{\av}[1]{\del{<}{>}{#1}}
\newcommand {\erf}{\mbox{Erf}}
\newcommand{\bi}{\begin{itemize}}
\newcommand{\ei}{\end{itemize}}
\newcommand {\vx}{\vec{x}}
\newcommand {\vy}{\vec{y}}
\newcommand {\vu}{\vec{u}}
\newcommand {\vb}{\vec{b}}
\newcommand {\vxi}{\vec{\xi}}
\newcommand {\vnabla}{\vec{\nabla}}
\newcommand {\bR}{{\bf R}}
\newcommand {\bnu}{{\bm \nu}}
\newcommand {\bnabla}{{\bm \nabla}}

\title{A linear theory for control of non-linear stochastic systems}
\author{Hilbert J. Kappen}
 \email{B.Kappen@science.ru.nl}
\affiliation{%
Department of Medical Physics \& Biophysics\\
Radboud University, Geert Grooteplein 21\\
6525 EZ Nijmegen, The Netherlands
}%

 \homepage{http://www.snn.kun.nl/~bert}

\date{\today}

\begin{abstract}
We address the role
of noise and the issue of efficient computation 
in stochastic optimal control problems. We consider a class of non-linear
control problems that can be formulated as a path integral
and where the noise plays the role of temperature.  The path integral
displays symmetry breaking and there exist a critical noise value that
separates regimes where optimal control yields qualitatively different
solutions. The path integral can be computed efficiently by Monte Carlo
integration or by Laplace approximation, and can therefore be used to
solve high dimensional stochastic control problems.

\end{abstract}

\maketitle

Optimal control of non-linear systems in the presence of noise 
is a very general problem that occurs in many areas of science and
engineering. It underlies autonomous system behavior, such as the
control of movement and planning of actions of animals and robots, but
also for instance the optimization of financial investment policies and
control of chemical plants.  The problem is simply stated: given that
the system is in this configuration at this time, what is the optimal
course of action to reach a goal state at some future time. The cost of
each time course of actions consists typically of a path contribution,
that specifies the amount of work or other cost of the trajectory,
and an end cost, that specifies to what extend the trajectory reaches
the goal state.

In the absence of noise, the optimal control problem can be solved in two
ways: using the Pontryagin Minimum Principle (PMP) \cite{pontryagin62}
which is a pair of ordinary differential equations that are similar to
the Hamilton equations of motion or using the Hamilton-Jacobi-Bellman (HJB)
equation, which is a partial differential equation \cite{bellman64}.

In the presence of (Wiener) noise, the PMP formalism is replaced by a set
of stochastic differential equations which become difficult to solve
(see however \cite{yong_zhou99}).  The inclusion of noise in the HJB
framework is mathematically quite straight-forward, yielding the so-called
stochastic HJB equation \cite{stengel93}.  Its solution
, however, requires a discretization of space and time and the computation
becomes intractable in both memory requirement and CPU time
in high dimensions.  As a result, deterministic control can be computed
efficiently using the PMP approach, but stochastic control is intractable
due to the curse of dimensionality.

For small noise, one expects that optimal stochastic control resembles
optimal deterministic control, but for larger noise, the optimal
stochastic control can be entirely different from the deterministic
control \cite{russell_norvig03}, but there is currently no good
understanding how noise affects optimal control.

In this paper, we address both the issue of efficient computation and
the role of noise in stochastic optimal control. We consider a class
of non-linear stochastic control problems, that can be formulated as
a statistical mechanics problem.
This class of control problems 
includes arbitrary
dynamical systems, but with a limited control mechanism. 
It contains linear-quadratic \cite{stengel93} control as a special case. 
We show that under certain conditions on the noise, the HJB equation
can be written as a {\em linear} partial differential equation
\be
-\partial_t \psi= H \psi
\label{linear_hjb}
\ee
with $H$ a (non-Hermitian) operator. Eq.~\ref{linear_hjb} must be
solved subject to a boundary condition at the end time.
As a result of the linearity of Eq.~\ref{linear_hjb},
the solution can be obtained in terms of a diffusion process evolving forward in time,
and can be written as a path integral.
The path integral has a direct interpretation as a
free energy, where noise plays the role of temperature.

This link between stochastic optimal control and a free energy has two
immediate consequences. 1) Phenomena that allow for a free energy description,
typically display phase transitions. We argue that for stochastic optimal
control one can identify a critical noise value that
separates regimes where the optimal control is qualitatively
different and illustrate this with a simple example.
2) Since the path integral appears in other
branches of physics, such as statistical mechanics and quantum
mechanics, we can borrow approximation methods from those fields to
compute the optimal control approximately. We show how the Laplace
approximation can be combined with Monte Carlo sampling to efficiently
compute the optimal control. 

Let $\vx$ be an $n$-dimensional stochastic variable that is subject to the
stochastic differential equation
\be
d\vx=(\vb(\vx,t) + \vu)dt + d\vxi
\label{easy_dynamics}
\ee
with $d\vxi$ a Wiener process with $\av{d\xi_i d\xi_j}=\nu_{ij}dt$, and
$\nu_{ij}$ independent of $\vx, \vu, t$. 
$\vb(\vx,t)$ is an arbitrary $n$-dimensional
function of $\vx$ and $t$, 
and $\vu$ an $n$-dimensional vector of control variables.
Given $\vx$ at an initial time $t$, the
stochastic optimal control problem is to find the control path
$\vu(\cdot)$ that minimizes
\bea
&&C(\vx,t,\vu(\cdot))=\nonumber\\
&&\av{\phi(\vx(t_f))+\int_{t}^{t_f}d \tau\left(\frac{1}{2}\vu(\tau)^T
\bR \vu(\tau)  + V(\vx(\tau),\tau)\right)}_{\vx}\nonumber\\
\label{easy_cost}
\eea
with $\bR$ a matrix, $V(\vx,t)$ a time-dependent potential, and $\phi(\vx)$
the end cost. The brackets
$\av{}_{\vx}$ denote expectation value with respect to the stochastic
trajectories~(\ref{easy_dynamics}) that start at $\vx$.

One defines the {\em optimal cost-to-go function} from any
time $t$ and state $\vx$ as
\bea
J(\vx,t)&=&\min_{\vu(\cdot)}C(\vx,t,\vu(\cdot)).
\eea
$J$ satisfies the stochastic HJB equation which takes the form
\bea
-\partial_t J&=&\min_{\vu}\left( \frac{1}{2}\vu^T \bR \vu  + V+(\vb+\vu)^T\vnabla J+
\frac{1}{2}\mathrm{Tr}\left(\bnu \bnabla^2 J\right)\right)\nonumber\\
&=&-\frac{1}{2}(\vnabla J)^T \bR^{-1} \vnabla J +V
+\vb^T \vnabla J + \frac{1}{2}\mathrm{Tr}\left(\bnu \bnabla^2
J\right)\nonumber\\
\label{hjb}
\eea
with $\mathrm{Tr}(\bnu \bnabla^2 J)=\sum_{ij}\nu_{ij}\partial^2
J/\partial x_i\partial x_j$ and
\be
\vu=-\bR^{-1}\vnabla J(\vx,t)
\label{u}
\ee
the optimal control at $\vx,t$.
The HJB equation is non-linear in $J$ and must be solved with end boundary condition
$J(\vx,t_f)=\phi(\vx)$. 

Define $\psi(\vx,t)$ through 
\endnote{The log transform goes back to Schr\"odinger and was first used in control theory by
\cite{fleming78}.}
\be
J(\vx,t)=-\lambda \log \psi(\vx,t)
\label{log_transform}
\ee
 and
assume there exists a scalar $\lambda$ such that
\be
\lambda \delta_{ij}= (\bR \bnu)_{ij}
\label{noise_condition}
\ee
with $\delta_{ij}$ the Kronecker delta.
In the one dimensional case, such a $\lambda$ can always be found. In
the higher dimensional case, this restricts the matrices 
$\bR \propto \bnu^{-1}$
\endnote{
For example, if 
both $\bR$ and $\bnu$ are
diagonal matrices,  in a direction with low noise, 
control is expensive ($R_{ii}$ large) and only small control steps
are permitted. In noisy directions the reverse is true: control is cheap
and large control values are permitted.
As another example, consider a one-dimensional 
second order system subject to additive control $\ddot{x}=b(x,t)+u$.
The stochastic formulation is of the form
\[
dx=y dt,\qquad dy=(b(x,t)+u)dt + d\vxi
\]
Eq.~\ref{noise_condition} states that due to 
the absence of a control term in the equation for $dx$, the noise in
this equation should be zero.}.
Eq.~\ref{noise_condition} 
reduces the dependence of optimal control on the $n$-dimensional noise matrix to a scalar value $\lambda$
that will play the role of temperature.
Eq.~\ref{hjb} reduces to the linear equation~\ref{linear_hjb} with
\bea
H&=&-\frac{V}{\lambda}
+\vb^T \vnabla + \frac{1}{2}\mathrm{Tr}(\bnu\bnabla^2)
\eea

Let $\rho(\vy,\tau|\vx,t)$ with $\rho(\vy,t|\vx,t)=\delta(\vy-\vx)$ describe a
diffusion process for $\tau>t$ defined by the Fokker-Planck equation
\be
\partial_\tau \rho= H^\dagger \rho = -\frac{V}{\lambda}\rho
-\vnabla^T (\vb\rho)
+ \frac{1}{2}\mathrm{Tr}(\bnu\bnabla^2) 
\rho
\label{fp}
\ee
with $H^\dagger$ the Hermitian conjugate of $H$. Then $A(\tau)=\int
d\vy
\rho(\vy,\tau|\vx,t) \psi(\vy,\tau)$ is independent of $\tau$ and in
particular
$A(t)=A(t_f)$. It immediately follows that
\bea
\psi(\vx,t)&=&
\int d\vy \rho(\vy,t_f|\vx,t) \exp(-\phi(\vy)/\lambda) 
\label{forward}
\eea
We arrive at the important conclusion that $\psi(\vx,t)$ can be
computed either by backward integration using
Eq.~\ref{linear_hjb} or by forward
integration of a diffusion process given by Eq.~\ref{fp}.

We can write the integral in Eq.~\ref{forward} 
as a path integral. We use the standard argument
\cite{kleinert95}
and divide the time interval $t\rightarrow t_f$ in $n_1$ intervals and
write $\rho(\vy,t_f|\vx,t)=\prod_{i=1}^{n_1}
\rho(\vx_i,t_{i}|\vx_{i-1},t_{i-1})$
and let $n_1 \rightarrow \infty$. The result is
\bea
\psi(\vx,t) &=& 
\int [d\vx]_{\vx} \exp \left(-\frac{1}{\lambda}S(\vx(t\rightarrow
t_f))\right)\label{path_integral}
\eea
with
$\int [d\vx]_{\vx}$ an integral over all paths $\vx(t\rightarrow t_f)$ that
start at $\vx$ and with
\begin{widetext}
\bea
S(\vx(t\rightarrow t_f))&=&\phi(\vx(t_f)+\int_t^{t_f}
d\tau\left(\frac{1}{2}\left(\frac{d\vx(\tau)}{d\tau}-\vb(\vx(\tau),\tau)\right)^T
\bR \left(\frac{d\vx(\tau)}{d\tau}-\vb(\vx(\tau),\tau)\right)+V(\vx(\tau),\tau)\right)
\label{path_action}
\eea
\end{widetext}
the Action associated with a path. From Eqs.~\ref{log_transform}
and~\ref{path_integral}, 
the cost-to-go $J(x,t)$ becomes a log partition sum (ie. a free energy) with temperature $\lambda$.

The path integral Eq.~\ref{path_integral} 
can be estimated by stochastic integration from $t$ to
$t_f$ of the
diffusion process Eq.~\ref{fp} in which particles get annihilated at a rate
$V(\vx,t)/\lambda$:
\bea
\vx&=&\vx+\vb(\vx,t) dt + d\vxi, \quad \mbox{with probability}~1-Vdt/\lambda\nonumber\\
\vx&=&\dagger, \quad \mbox{with probability}~Vdt/\lambda
\label{diffusion}
\eea
where $\dagger$ denotes that the particle is taken out of the
simulation.
Denote the trajectories 
by $\vx_\alpha(t\rightarrow t_f),\alpha=1,\ldots,N$. Then,
$\psi(\vx,t)$ and $\vu$ are 
estimated as
\bea
\hat{\psi}(\vx,t)&=& \sum_{\alpha\in \mathrm{alive}}w_\alpha
\label{mc_psi}\\
\hat{\vu}dt&=&\frac{1}{\hat{\psi}(\vx,t)}\sum_{\alpha\in \mathrm{alive}}^N
w_\alpha d\vxi_\alpha(t)\label{mc_u}\\
w_\alpha&=&\frac{1}{N} \exp(-\phi(\vx_\alpha(t_f))/\lambda)\nonumber
\eea
where 'alive' denotes the subset of trajectories that do not get killed
along the way by the $\dagger$ operation. The normalization $1/N$
ensures that the annihilation process is properly taken into account.
Eq.~\ref{mc_u} states that optimal
control at time $t$ is obtained by averaging the initial directions
of the noise component of the trajectories
$d\vxi_\alpha(t)$, weighted by their success at $t_f$.

The above sampling procedure can be quite inefficient, when many
trajectories get annihilated.  One of the simplest procedures to improve
it is by importance sampling.  We replace the diffusion process that
yields $\rho(\vy,t_f|\vx,t)$ by another diffusion process, that will yield
$\rho'(\vy,t_f|\vx,t)=\exp(-S'/\lambda)$.  Then Eq.~\ref{path_integral}
becomes,
\beaa
\psi(\vx,t)&=& \int [d\vx]_{\vx} \exp\left(-S'/\lambda\right)
\exp\left(-(S-S')/\lambda\right) \eeaa

The idea is to chose $\rho'$ 
such as to make the sampling of the path integral as
efficient as possible. 
Here, we use the Laplace approximation,
which is given by the $k$ deterministic trajectories 
$x_\beta(t\rightarrow t_f)$
that
minimize the Action
\bea
J(\vx,t)&\approx&
-\lambda\log \sum_{\beta=1}^k \exp(-S(\vx_\beta(t\rightarrow
t_f)/\lambda)
\label{laplace_multi}
\eea
The Laplace approximation ignores all fluctuations around the modes
and becomes exact in the limit
$\lambda\rightarrow 0$.
The Laplace approximation can be computed efficiently, requiring ${\cal O}(n^2 m^2)$
operations, where $m$ is the number of time discretization. 

For each Laplace trajectory, we define a diffusion processes
$\rho'_\beta$ according to Eq.~\ref{diffusion}
with $\vb(\vx,t)=\dot{\vx}_\beta(t)$.
The estimators for $\psi$ and $\vu$ are given again by 
Eqs.~\ref{mc_psi} and~\ref{mc_u}, but with weights
\bea
w_\alpha&=&\frac{1}{N}\exp
\left(-\left( S(\vx_\alpha(t\rightarrow t_f))-S'_\beta(\vx_\alpha(t\rightarrow
t_f))\right)/\lambda\right).\nonumber\\
\label{importance_weights}
\eea
$S$ is the original Action Eq.~\ref{path_action} and $S'_\beta$ 
is the new Action for the Laplace guided diffusion.
When there are multiple Laplace trajectories one should include all of
these in the sample.

\label{section:slit}
We give a simple one-dimensional example of a double slit to illustrate
the effectiveness of the Laplace guided MC method and to show how the
optimal cost-to-go undergoes symmetry breaking as a function of the
noise.

\begin{figure}
\bc
\includegraphics[angle=0, height=0.2\textwidth]{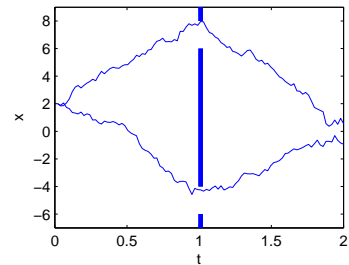}
\ec
\caption{ 
A double slit is placed at $t=1$ with openings at $-6<x<-4$ and
$6<x<8$. $V=\infty$ for $t=1$ outside the openings, and zero otherwise.
Also shown are two example trajectories under optimal control. 
} \label{slit1} \end{figure}

\begin{figure}
\bc
\includegraphics[angle=0, height=0.2\textwidth]{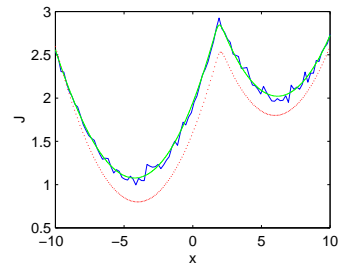}
\ec
\caption{ 
Comparison of Laplace approximation (dotted line) and Monte Carlo
importance sampling (solid jagged line) of $J(x,t=0)$ with exact result
(solid smooth line) for the double slit problem. The
importance sampler used $N=100$ trajectories for each $x$.  $R=0.1,
\nu=1, dt=0.02$.  } \label{slit2} \end{figure}

Consider a stochastic particle that moves with constant
velocity from $t=0$ to $t_f=2$ in the horizontal direction and where
there is deflecting noise in the $x$ direction:
\beaa
dx &=& u dt + d\xi
\eeaa
The cost is given by Eq.~\ref{easy_cost} with
$\phi(x)=\frac{1}{2} x^2$ and
$V(x,t_1)$ implements a slit at an intermediate time $t_1=1$
(Fig.~\ref{slit1}).
Solving the cost-to-go by means of the forward computation using
Eq.~\ref{forward} can be done in closed form. The exact result,
the Laplace approximation Eq.~\ref{laplace_multi} and the Laplace guided importance sampling
result using Eq.~\ref{importance_weights} are plotted for $t=0$ as a
function of $x$ in Fig.~\ref{slit2}.
For each $x$, the Laplace approximation consists of the two deterministic
trajectories, each being piecewise linear, starting at $t=0$ in $x$
and ending at $t=2$ in $x=0$.
We see that the Laplace approximation is quite good for this example, in 
particular when one takes into account that a constant shift in $J$ does
not affect the optimal control. The MC importance sampler has maximal
error of order 0.1 and is significantly better
than the Laplace approximation.
Naive MC sampling using Eq.~\ref{diffusion} (not shown) fails for this problem, because most
trajectories get killed by the infinite potential.  Numerical simulations
using $N=100000$ trajectories yield estimation errors in $J$ up to
approximately 6 for certain values of $x$.

We show an example how optimal stochastic control exhibits spontaneous
symmetry breaking. For two slits of width $\epsilon$ at $x=\pm 1$, the
cost-to-go  
becomes to lowest order in $\epsilon$:
\beaa
J(x,t)&=&
\frac{R}{T}\left(\frac{1}{2}x^2-\nu T \log 2 \cosh\frac{x}{\nu
T}\right)+\mathrm{const.}, \quad t< t_1
\eeaa
where the constant diverges as ${\cal O}(\log \epsilon)$ independent
of $x$ and $T=t_1-t$ the time to reach the slits.
The expression between brackets is a typical free energy with inverse temperature $\beta=1/\nu T$. 
It displays a symmetry breaking at $\nu T= 1$.
The optimal control is given by the gradient of $J$:
\bea
u=\frac{1}{T}\left(\tanh \frac{x}{\nu T} -x\right)
\label{u_delayed}
\eea
For $T>1/\nu$ (far in the past) optimal control steers towards $x=0$ (between the targets)
and delays the choice which slit to aim for until later.  The reason why
this is optimal is that the expected diffusion alone
of size $\sqrt{\nu T}$ is likely to reach any of the slits without control
(although it is not clear yet which slit).
Only sufficiently late in time ($T<1/\nu$) should one make
a choice.  
\begin{figure}
\bc
\includegraphics[angle=0, height=0.3\textwidth]{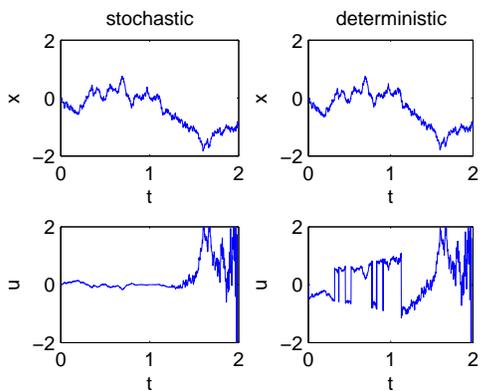}
\ec
\caption{
Symmetry breaking in $J$ as a function of $T$ implies a 'delayed
choice'
mechanism for optimal stochastic control.
When the target is far in the future,
the optimal policy is to steer between the targets. Only when $T<1/\nu$
should one aim for one of the targets.
Sample trajectories (top row) and controls (bottom row) 
under stochastic control
(left column) and deterministic control
(right column). $\nu=R=1$, $t_1=2$.} \label{multi_laplace}
\end{figure}

Figure~\ref{multi_laplace} depicts two trajectories and their
controls under stochastic optimal control (Eq.~\ref{u_delayed}) 
and deterministic optimal control (Eq.~\ref{u_delayed} with $\nu=0$), using the
same realization of the noise. 
Note, that at early times
the deterministic control drives $x$ away
from zero whereas in the
stochastic control drives $x$ towards zero and is smaller in size. 
The stochastic control delays the choice
for which slit to aim until $T\approx 1$.

In summary, we have shown that stochastic optimal control involves
symmetry breaking with qualitatively different solutions for high and low
noise levels. This property is expected to be true also for more general
stochastic control problems. The path integral formulation allows for an
efficient solution of the HJB equation because it replaces the intractable $n$-dimensional
numerical integration by a Monte Carlo sampling, which is known to be
often
much more efficient.  
This approach will thus be of direct practical value for
the control of high dimensional, strongly non-linear, systems, such as 
for instance
robot arms, navigation of autonomous systems, and chemical reactions.  
 For realistic
applications, naive sampling should be replaced by more advanced sampling
schemes, such as importance sampling or a Metropolis method, and should
be combined with efficient discretization such as splines, wavelets
or a Fourier basis \cite{miller75_jchemphys75,freeman_jchemphys81}.

\begin{acknowledgments}
I would like to thank Hans Maassen for useful discussions.  This work is
supported in part by the Dutch Technology Foundation and the BSIK/ICIS
project.  \end{acknowledgments}

\bibliography{/home/snn/bertk/doc/authors}
\end{document}